\documentclass[aps,prl,twocolumn,superscriptaddress,amsmath,amssymb]{revtex4}
\usepackage{graphicx}
\usepackage{epstopdf}
\usepackage{amsmath,amsfonts,amssymb,mathtools}
\usepackage{color,xcolor}
\usepackage{soul}
\usepackage{braket}
\usepackage[normalem]{ulem}
\usepackage{graphicx}
\usepackage{subfigure}
\usepackage{float}
\usepackage{mathrsfs}
\usepackage{bbold}
\usepackage{psfrag}
\usepackage{mathcomp}
\usepackage{verbatim}
\usepackage{multirow}
\usepackage{diagbox}
\usepackage[colorlinks,citecolor=blue]{hyperref}

\usepackage{simplewick}

\def\ba{\begin{eqnarray}}
\def\ea{\end{eqnarray}}

\begin{document}

\title{Detecting Fermi Surface Nesting Effect for Fermionic Dicke Transition by Trap Induced Localization}

\author{Shi Chen}
\affiliation{Graduate School of China Academy of Engineering Physics, Beijing, 100193, China}

\author{Yu Chen}
\email{ychen@gscaep.ac.cn}
\affiliation{Graduate School of China Academy of Engineering Physics, Beijing, 100193, China}
\date{\today}

\begin{abstract}
Recently, the statistical effect of fermionic superradiance is approved by series of experiments both in free space and in a cavity. The Pauli blocking effect can be visualized by a 1/2 scaling of Dicke transition critical pumping strength against particle number $N_{\rm at}$ for fermions in a trap. However, the Fermi surface nesting effect, which manifests the enhancement of superradiance by Fermi statistics is still very hard to be identified.  Here we studied the  influence of localized fermions on the trap edge when both pumping optical lattice and the trap are presented. We find due to localization, the statistical effect in superradiant transition is enhanced. Two new scalings of critical pumping strength are observed as $4/3$, and $2/3$ for mediate particle number, and the Pauli blocking scaling $1/3$ (2d case) in large particle number limit is unaffected. Further, we find the $4/3$ scaling is subject to a power law increasing with rising ratio between recoil energy and trap frequency in pumping laser direction $E_R/\omega_x$. The divergence of this scaling of critical pumping strength against $N_{\rm at}$ in $E_R/\omega_x\rightarrow\infty$ limit can be identified as the Fermi surface nesting effect. Thus we find a practical experimental scheme for visualizing the long-desired Fermi surface nesting effect with the help of trap induced localization in a two-dimensional Fermi gas in a cavity.  
\end{abstract}

\maketitle

{\color{blue}\emph{Introduction}}  In recent decades, the developments in achieving strong coupling between atoms and light have led us to a new platform for studying non-equilibrium open quantum systems\cite{Esslinger07,Colombe07}. With these accumulation, the Dicke model, a typical model of strong interactions between atoms and light is finally realized\cite{Esslinger10}. It ends the era for Dicke transition being purely theoretical\cite{Dicke54,Lieb73}. The Dicke transition manifests itself through the emergence of the steady state superradiance together with a checkerboard density order in atoms\cite{Esslinger11}. The spontaneity of the self-organized crystalline is verified by the roton mode softening \cite{Roton12} and the critical behavior of the dynamical structure factor\cite{Esslinger13,Hemmerich14,Esslinger15}. Exotic phases like density-ordered Mott insulators\cite{Hemmerich15ex,Esslinger16,Esslinger18,Esslinger20}, as well as supersolid breaking U(1) symmetry and translation symmetry \cite{ContinousSS17,EsslingerSS17,EsslingerSS18} are observed successively experimentally. Excited topics with the combinations of quantum many-body systems and the traditional presentation of an open quantum system --- cavity-QED (quantum electrodynamics) are enlightened following these new advances\cite{Review21}.

Besides the developments in the superradiance of Bose gases, there are also many interesting statistical effects in fermionic superradiance. The most prominent signature for Dicke transitions of degenerate Fermi gases is its density dependence, namely, the Fermi surface nesting effect and the Pauli blocking effect\cite{Simons14,Piazza14,Yu14}. The Fermi surface nesting effect is due to the resonance between pairs of nested states on the Fermi surface whose momentum difference matches the cavity photon momentum. It results a sharp decreasing of critical pumping strength at specific fillings. The Pauli blocking effect, on the other hand, leads to a suppression of superradiance due to both of the momentum states connected by photon absorption being occupied. There are also further effects like liquid-gas like a transition for p-band filling\cite{Simons14} and statistical crossover in interacting Fermi gases\cite{Yu15}, etc. For a long-time, these studies for statistical effects in fermionic superradiance are only theoretical. In a recent experiment, a steady state superradiance of fermions is realized in a cavity for the first time. A scaling law of critical pumping strength as $N^{-1/2}_{\rm at}$ against the particle number $N_{\rm at}$ at large $N_{\rm at}$ limit is verified in a three-dimensional trapped fermions system\cite{Wu21}. This is in sharp contrast with the bosonic Dicke transition whose scaling law is $N_{\rm at}^{-1}$. Pauli blocking effect in free space superradiance is also verified thereafter\cite{Ketterle21,Deb21,YeJ21}.  Unfortunately, contrary to the well-established Pauli blocking effect,  Fermi surface nesting effect, which shows the enhancement of superradiance by Fermi statistics is still not verified in experiment. The main difficulty is the presence of the trap makes density ill-defined.

\begin{figure}[t]
\includegraphics[width=8.5cm]{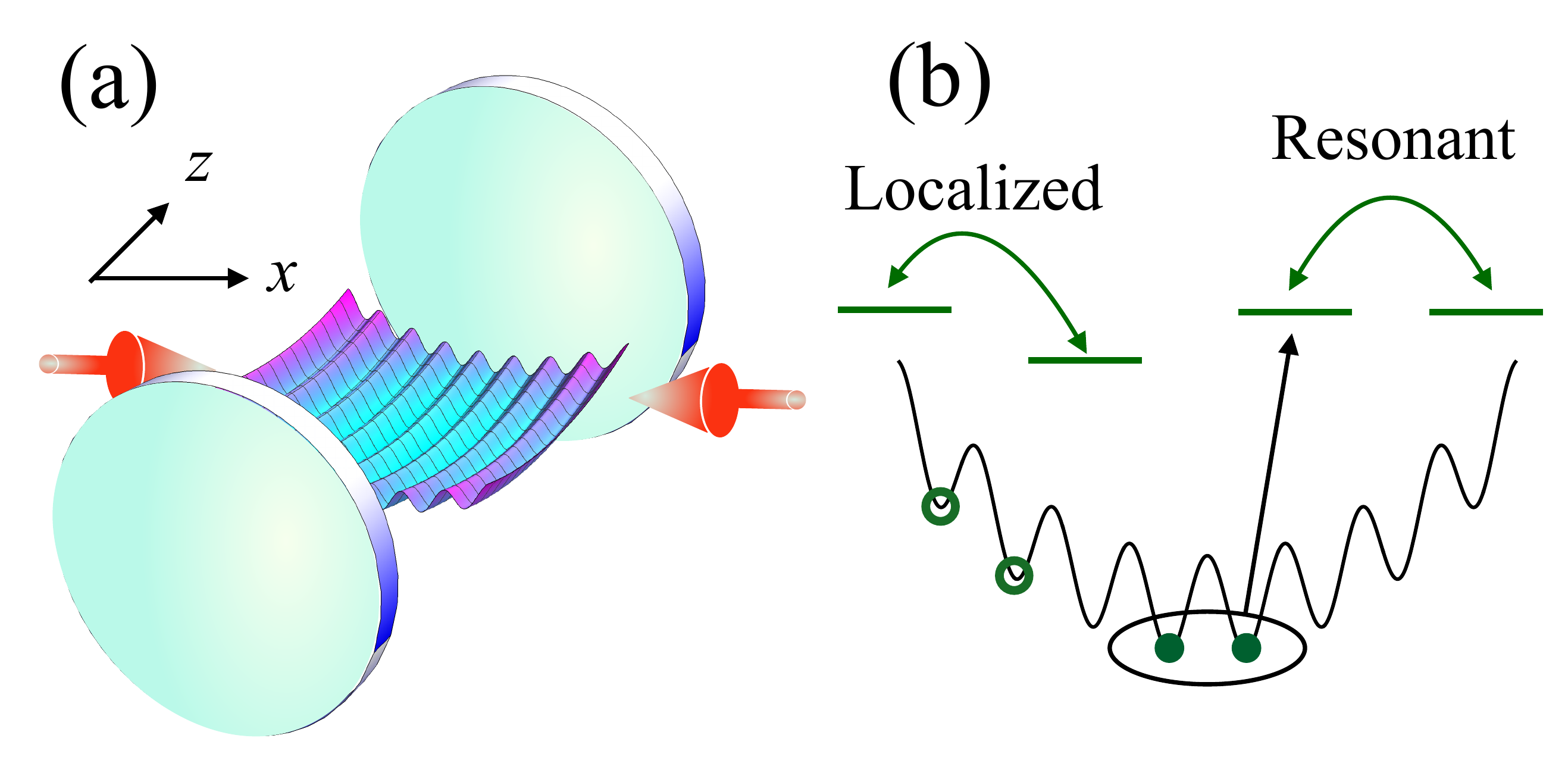}
\caption{ In (a), we show the illustration of our setup. Two-dimensional Fermi gases are put into a cavity with a harmonic trap in $x$ and $z$ directions. In (b), we show the mechanism for localization on the trap edge. At the bottom of the trap, fermions at neighbor sites are resonant, thus Bloch state is still a good approximation. On the edge of the trap, the on-site energy difference becomes larger than the hopping strength, which leads to localization. }
\label{Mechanism}
\end{figure}

In this Letter, we offer a method to identify the Fermi surface nesting effect for two-dimensional Fermi gases within a trap. We find that, when trap and optical lattice are both presented, there are localized states on the trap edge. The mechanism is shown in Fig.~\ref{Mechanism}(b). One can observe that on the trap edge, the onsite energy difference will outgo the hopping strength, which leads to localization. These localizations then contribute a lot of one-dimensional fermion tubes on the trap edge, where Fermi surface nesting effect is prominent. Phenomenologically, we find for different cavity detune $\Delta_c$, the critical pumping strength of Dicke transition as a function of particle number $N_{\rm at}$ falls into two universal functions up to a constant shift in log-log plot. One of the universal curves for small $\Delta_c$ shows a universal scaling $N_{\rm at}^{-1}$ for small $N_{\rm at}$ and $N_{\rm at}^{-1/3}$ at large $N_{\rm at}$ limit. There is no scaling faster than $N_{\rm at}^{-1}$ in this typical curve. Let us denote the scaling as $\varkappa$ for a critical pumping strength scaling law $N_{\rm at}^{-\varkappa}$ to simplify our description.  For larger $\Delta_c$, however, we find a new scaling $\varkappa>1$ ($\varkappa=1.33$) in intermediate particle number region. 
We also find a crossover between these two typical critical pumping strength functions, which is sharp within a very small window of $\Delta_c$. In the same crossover $\Delta_c$ region,  we find the eigenstate on the Fermi surface turns from an extended state to a localized state, which identifies the mechanism. Further, $\varkappa\rightarrow\infty$ in $\omega_z/E_R\rightarrow 0$ limit at a specific $N_{\rm at}$ is obtained as a result of Fermi surface nesting whose tendency is verified numerically. Therefore, the measurement of critical pumping strength as a function of particle number $N_{\rm at}$ for different cavity detune $\Delta_c$ and different trap frequencies can help us to identify Fermi surface nesting effect. We also stress that the trap induced localization is vital in detecting Fermi surface nesting effect, quite opposite from the original impression of trap being nothing but an obstacle.

{\color{blue}\emph{Model}} We consider spinless fermions trapped inside a high-Q single mode cavity within a harmonic trap($\hbar=1$ throughout),
\ba
\hat{H}&=&\hat{H}_{at}-\Delta_c\hat{a}^{\dag}\hat{a},\label{Eq1}
\ea
\ba
\hat{H}_{at}=\int d\textbf{r}\hat{\psi}^{\dag}(\textbf{r})(\hat{H}_{0}+\eta(\textbf{r})(\hat{a}^{\dag}+\hat{a})+U_0(\textbf{r})\hat{a}^{\dag}\hat{a})\hat{\psi}(\textbf{r}),
\ea
\ba
\hat{H}_{0}=\frac{-\nabla^2}{2m}+V_P(\textbf{r})+\frac{1}{2}m\omega^2\textbf{r}^2,
\ea
Here $\hat{a}$ is the annihilation operator of the cavity field, $\hat{\psi}$ is an annihilation operator of fermionic atoms. Here $V_P(\textbf{r})=V_P\cos^2(k_0x)$ is the optical lattice generated by the pumping field, $U_0(\textbf{r})=U_0\cos^2(k_0z)$ is the cavity field self-energy, $\eta(\textbf{r})=\eta_0\cos(k_0x)\cos(k_0z)$ is the interference between the pumping field and the cavity field. To be more specific, $V_P=\Omega^2_p/\Delta_a$, $U_0=g_0^2/\Delta_a$,  and $\eta_0=g\Omega_p/\Delta_a$, where $\Delta_a=\omega_a-\omega_p$ is AC stark shift ($\omega_a$ is the excited state energy), $\Delta_c=\omega_p-\omega_c$ is the cavity detuning. Here the cavity in consideration has a photon decay rate of $\kappa$. $\Omega_p$ is the strength of the pumping lasers, $g_0$ is the single-photon Rabi frequency of the cavity field, and $k_0$ is the wave number of the pumping field which is close to the wave number of the cavity field. The wave number of the cavity field is  approximated by $k_0$. The recoil energy is defined by $E_R\equiv k_0^2/2m$.

{\color{blue}\emph{Mean Field Theory}} Since the cavity photon is lossy, the equation of cavity photon field should follow Lindblad equation,
\ba
i\partial_t\hat{a}=[\hat{a},\hat{H}]+2\kappa{\cal L}_{\hat{a}}\hat{a},
\ea
where ${\cal L}_{\hat{a}}\hat{a}=\hat{a}^\dag \hat{a}\hat{a}-\frac{1}{2}\{ \hat{a}^\dag \hat{a},\hat{a}\}$ is the Lindblad operator on cavity field operator $\hat{a}$. Assuming that the cavity field is in a coherent state $|\alpha\rangle$, such that $\langle\alpha|\hat{a}|\alpha\rangle=\alpha$. Then the above equation can be written as
\begin{equation}
\begin{aligned}
\partial_t\alpha&=i(-\eta_0\Theta+(\Delta_c'+i\kappa)\alpha),
\end{aligned}
\label{Eq4}
\end{equation}
where the effective detuning $\Delta_c'(\alpha)=\Delta_c-\int d\textbf{r}U(\textbf{r})n(\textbf{r})$ and $\Theta(\alpha)=\int d \textbf{r}n(\textbf{r})\eta(\textbf{r})/\eta_0$. The
fermionic density function is $n(r)\equiv\braket{\hat{\psi}^{\dag}(\textbf{r})\hat{\psi}(\textbf{\textbf{r}})}={\rm Tr} (\hat{\psi}^{\dag}(\textbf{r})\hat{\psi}(\textbf{\textbf{r}})\hat{\rho}(t))$. Here ${\rm Tr}$ is over the atom's Hilbert space and a coherent state of cavity field is assumed. Here we assume the steady state density matrix of atoms is $\hat{\rho}_{\rm st}=e^{-\beta\hat{H}_{at}(\alpha)}/Z$ ($Z={\rm Tr}(e^{-\beta\hat{H}_{at}(\alpha)})$), which is justified by a full dynamical handling by Keldysh contour method. Here, $\hat{H}_{at}(\alpha)$ is defined by replace $\hat{a}$ in $\hat{H}_{at}$ with $\alpha$. In the steady state, $\partial\alpha/\partial t=0$, we have
\ba \alpha=\frac{\eta_0\Theta(\alpha)}{\Delta_c'(\alpha)+i\kappa},\label{Eq5}
\ea
As $\alpha$ is complex, the above steady state equation is indeed two equations. If we assume the Dicke transition is a second order transition, then ${\cal B}=\int d{\bf r}U({\bf r})n({\bf r})\approx U_0N/2$. The phase of $\alpha$ is locked by $\kappa$. The other equation can be understood as a minimization of free energy ${\cal F}_\alpha$,
\ba 
{\cal F}_\alpha\equiv -\beta^{-1}\log {\rm Tr}(e^{-\beta\hat{H}_{at}(\alpha)}).
\ea 
One can check $\eta_0\Theta+\alpha\int d{\bf r}U({\bf r})n({\bf r})=\partial {\cal F}_\alpha/\partial \alpha^*$. Then the steady state equation become,
\ba
\partial {\cal F}_\alpha/\partial \alpha^*-\Delta_c\alpha=0.
\ea

Further we can find in the second-order expansion of ${\cal F}_\alpha$, ${\cal F}_\alpha=-\eta^2_0\chi(\alpha+\alpha^*)^2+{\cal B}\alpha^*\alpha$. $\chi=\int d{\bf r}d{\bf r'}\langle \delta n({\bf r})\eta({\bf r})\delta n({\bf r'})\eta({\bf r}')\rangle/\eta^2_0$ is the static structure factor, which characterizes the density fluctuation at momentum $\pm k_0{\bf e}_x+\pm k_0{\bf e}_z$.
\ba
\chi=\frac{1}{2\eta^2_0}\sum_{n,n'}\left| \int d \textbf{r}\phi^*_n(\textbf{r})\phi_{n'}(\textbf{r})\eta(\textbf{r})\right|^2\frac{n_F(E_n)-n_F(E_n')}{E_{n'}-E_n}\label{StaticStructure}.
\ea 
Here $\phi_n({\bf r})$ is the eigenstate of $\hat{H}_{\rm 0}$, $E_n$ is the corresponding eigen energy of state $|n\rangle$,  $n_F$ the Fermi distribution function. If we assume the Dicke transition is second order transition, the critical pumping strength is obtained on condition that $\Delta'_c+4\eta^2_0\chi \Delta^{\prime 2}_c/(\Delta^{\prime 2}_c+\kappa^2)=0$ for the second order coefficient of $\alpha$ in ${\cal F}_\alpha$ being zero. The critical pumping strength is
\ba
\left| \frac{V_P^{cr}(N_{at})}{E_R} \left|=\frac{-(\Delta_c'^2+\kappa^2)E_R}{4U_0\Delta_c'\chi(N_{\rm at})}\right.\right..\label{Cr}
\ea
In this Letter, we will focus on the scaling of $V_{\rm P}^{\rm cr}$ over $N_{\rm at}$ to identify the statistical effect in fermionic superradiance transition in zero temperature.
\begin{figure}[t]
\includegraphics[width=8cm]{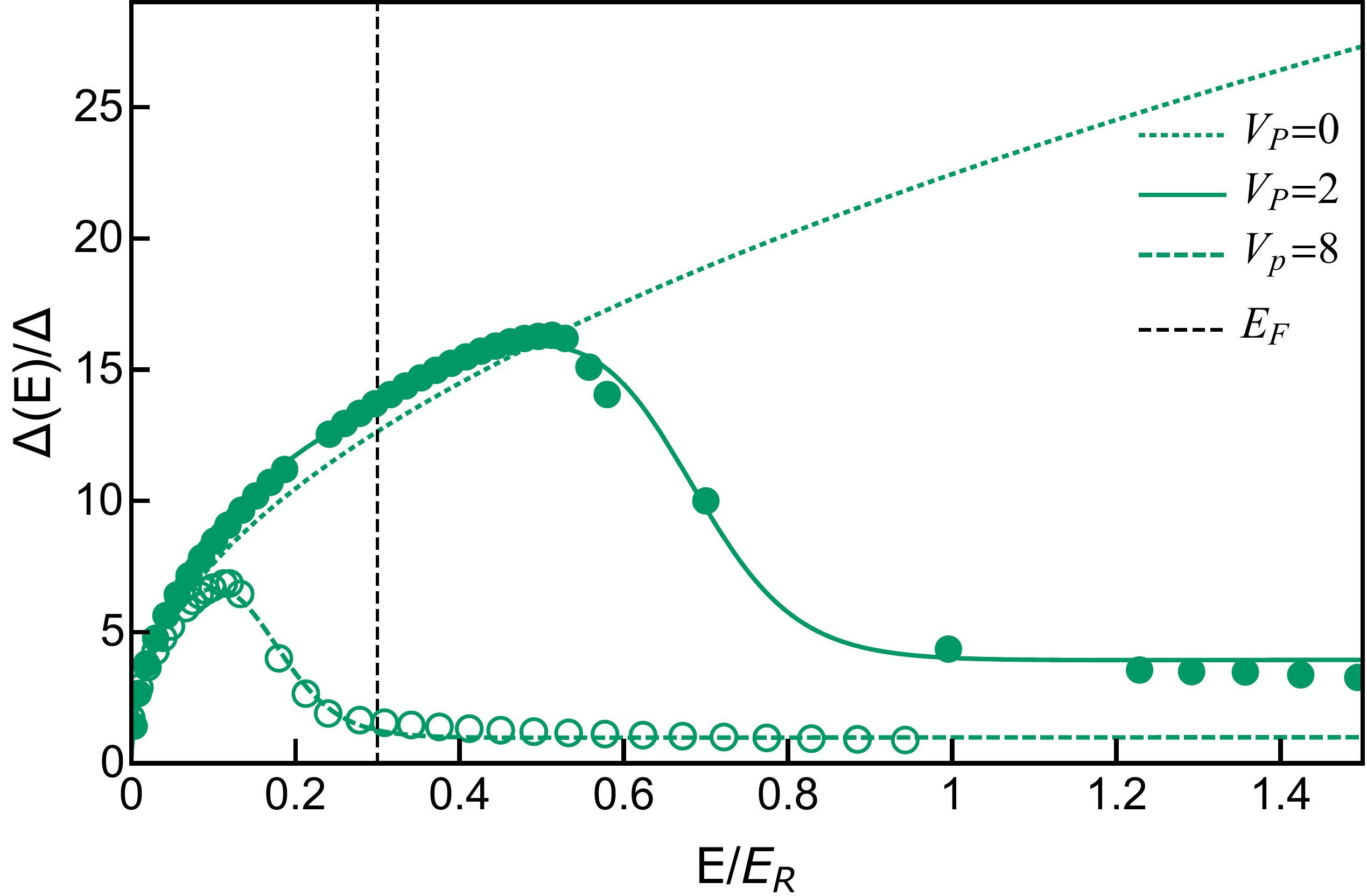}

\caption{The width $\Delta(E)$ of different eigenstates of $\hat{H}_y$ for different $V_0$. $\Delta=\frac{\pi}{k_0}$ is lattice spacing. Black dashed line represent typical fermi energy in our set-up. }
\label{a}
\end{figure}

{\color{blue}\emph{Localization on Trap Edge}} In previous calculation for fermions in a trap, we have assumed that the fermions states are more close to plane wave. However, on the edge of the trap, the interplay between the pumping lattice and the trap can lead to localization. Here we show the mechanism in a two-dimensional non-interacting fermions in a two-dimensional trap. 
\ba
\hat{H}_{\rm 0}=-\frac{\partial_x^2}{2m}+V_P(x)+\frac{m\omega_0^2}{2}x^2-\frac{\partial_z^2}{2m}+\frac{m\omega_0^2}{2}z^2
\ea
The eigenstate $|n\rangle$ can be factorized as $\phi_{n_x}(x)\phi_{n_z}(z)$. $\phi_{n_z}(z)$ is the eigenstate of Harmonic trap, and its eigen-energy is $n_z\omega_0$. $\hbar=1$.  Further, we solve the x direction hamiltonian with the trap, with $u_{1,k_x}(x)$ being the lowest band Bloch wave function. Then we can construct a Wannier basis by
\begin{equation}
    w_j(x)=\int dk_x e^{-ik_xjd}u_{1,k_x}(x),
\end{equation}
where $j$ is the site index, $d=\frac{\pi}{k_0}$. In the Wannier basis, the hamiltonian in x direction can be rewritten as
\begin{equation}
\begin{split}
 \hat{H}_x&=-\frac{\partial_x^2}{2m}+\frac{1}{2}m\omega^2x^2+V_P\cos^2(k_0x)
        \\&=\sum_j \mu_j\ket{j}\bra{j}+t\ket{j}\bra{j+1}+t\ket{j}\bra{j-1},
\end{split}   
\end{equation}
where, $t=\int dxw_i(x)(-\frac{\partial^2_x}{2m}+V_p\cos^2(k_0x))w^*_{i+1}(x)$, $\mu_i=\int dy|w_i(x)|^2\frac{1}{2}\omega^2x^2$. Next-to-nearest neighbor hopping can be also included, which is not explicitly shown in above equation.

One can check that when $|\mu_{j+1}-\mu_j|> t$, the eigen-state is localized because the resonance condition between sites is broken down.  Here we employ the wave packet width to quantitatively characterize the localization degree of these eigenstates for different eigen-energy. The width of a state is defined as: 
\ba
\Delta(E_n)=\sqrt{\langle x^2\rangle_n-\langle x\rangle_n^2},
\ea
where $\langle x^2\rangle_n=\int dx x^2|\phi_n(x)|^2$, $\langle x\rangle_n=\int dx x|\phi_n(x)|^2$. In Fig.\ref{a}, we show $\Delta(E_n)$ as a function of excitation energy $E_n$ for different pumping lattice strengths for fixed trap frequency. A clear sign for mobility edge is shown, and the high energy states can not be approximated as extended states. In the following, we will explore the physical consequences by the trap edge localization. 

\begin{figure}[b]
\includegraphics[width=8.5cm]{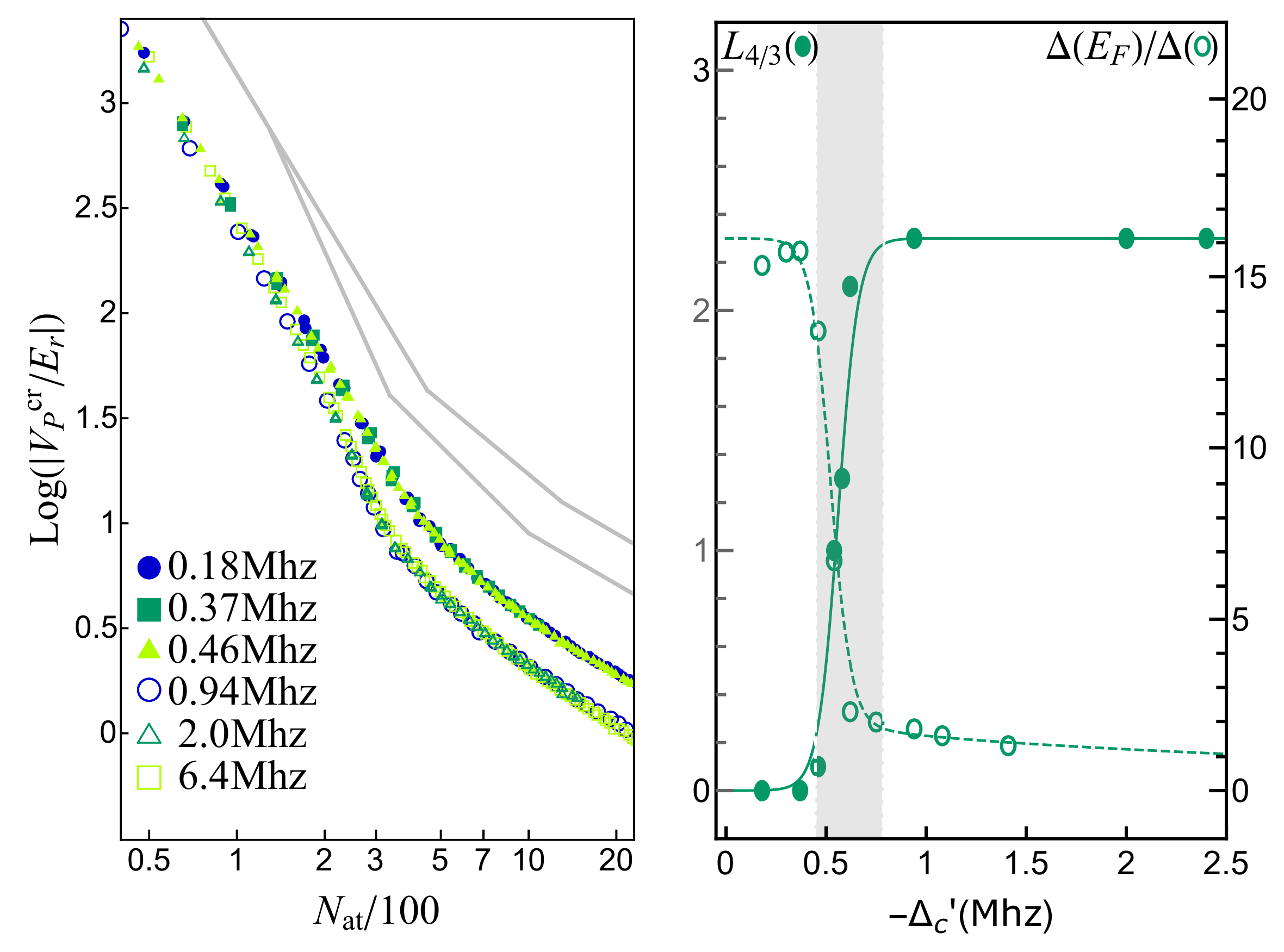}
\caption{ We fix $\kappa=1MHz, U_0=-100Hz$. In (a), we show the critical pumping strength of the Dicke transition as a function of $N_{\rm at}$ for different $\Delta_c$. In (b), the dashed line show the packet width of state near fermi surface, solid line represent the length with a gradient of 1.33 in fig (a). }
\label{UniversalScaling}
\end{figure}

\color{blue}\emph{Scaling of the Critical Pumping Strength}\color{black}   --- Following Eq.~\ref{StaticStructure} and Eq.~\ref{Cr}, together with the calculation of eigen-state $|n\rangle=|n_x,n_z\rangle$, we can obtain the numerical result for $V_P^{\rm cr}$ for different $\Delta_c$ and particle number $N_{\rm at}$. Here we have fixed $\omega_x=\omega_z=E_R/50$. One can find when $\Delta'_c$ is larger, the critical pumping strength is larger, thus the localization effect at the trap edge is larger. The numerical result for critical pumping strength of particle number $N_{\rm at}$ for different $\Delta'_c$ is given in Fig.~\ref{UniversalScaling}(a). One can find for smaller $\Delta'_c$, the log-log plot of $V_P^{\rm cr}(N_{\rm at})$ up to a multiplier constant factor falls into one universal curve. The initial scaling of $V_P^{\rm cr}$ against $N_{\rm at}$ is $-1$, then it crosses over to $-0.66$ and finally it crosses over to $-0.35$ in large $N_{\rm at}$ limit. The scaling reduction is due to Pauli blocking effect. One can also find there are no scalings larger than $1$, which means we only observe the suppression of superradiance due to Fermi statistics. The situation is suddenly changed when $\Delta'_c>0.5$MHz. We find due to localization on the trap edge, the critical pumping strength function  $V_P^{\rm cr}(N_{\rm at})$ fall into a new universal curve.  In this new universal curve, the initial small $N_{\rm at}$ scaling and the large particle number scaling are the same as the previous universal curve. What is beyond, a new scaling being larger than $1$ emerges around the middle range $N_{\rm at}$. This middle range $N_{\rm at}$ matches the fermion density for Fermi surface nesting effect. We compared the crossover region of the universal critical pumping strength curve and localization effect shown by the typical width of wave function $\phi_{n_x}(x)$ at the Fermi level. We find these two regions coincide with each other. Here we define $L_{1.33}$ as the length of $1.33$ scaling in $V_P^{\rm cr}(N_{\rm at})$. Therefore we conclude that the Pauli blocking scaling $0.35$ is not affected by localization and a new scaling $1.33$ emerges due to localization effect. Thus we find the evidence of Fermi surface nesting effect can be magnified by localization effect induced by the trap. 

To go further, we will now present a theoretical effective theory to understand why the Pauli blocking effect is immune to the localization effect in large $N_{\rm at}$ limit and whether the new scaling $1.33$ can be interpreted as the Fermi surface nesting effect. Before we give our analysis, we will first give our conclusion. The predictions of our effective theory are 1) the large $N_{\rm at}$ limit of $V_P^{\rm cr}$ is $V_P^{\rm cr}\propto N^{-1/3}$ and it is not changed by localization effect; 2) there is a kink at some middle range $N_{\rm at}$, and the critical pumping strength scaling of $N_{\rm at}$ becomes divergent before the kink in $\omega_z\rightarrow 0$ limit. The second phenomenon is a signature of Fermi surface nesting effect. 

\begin{figure}[t]
\hspace{-2ex}\includegraphics[width=8cm]{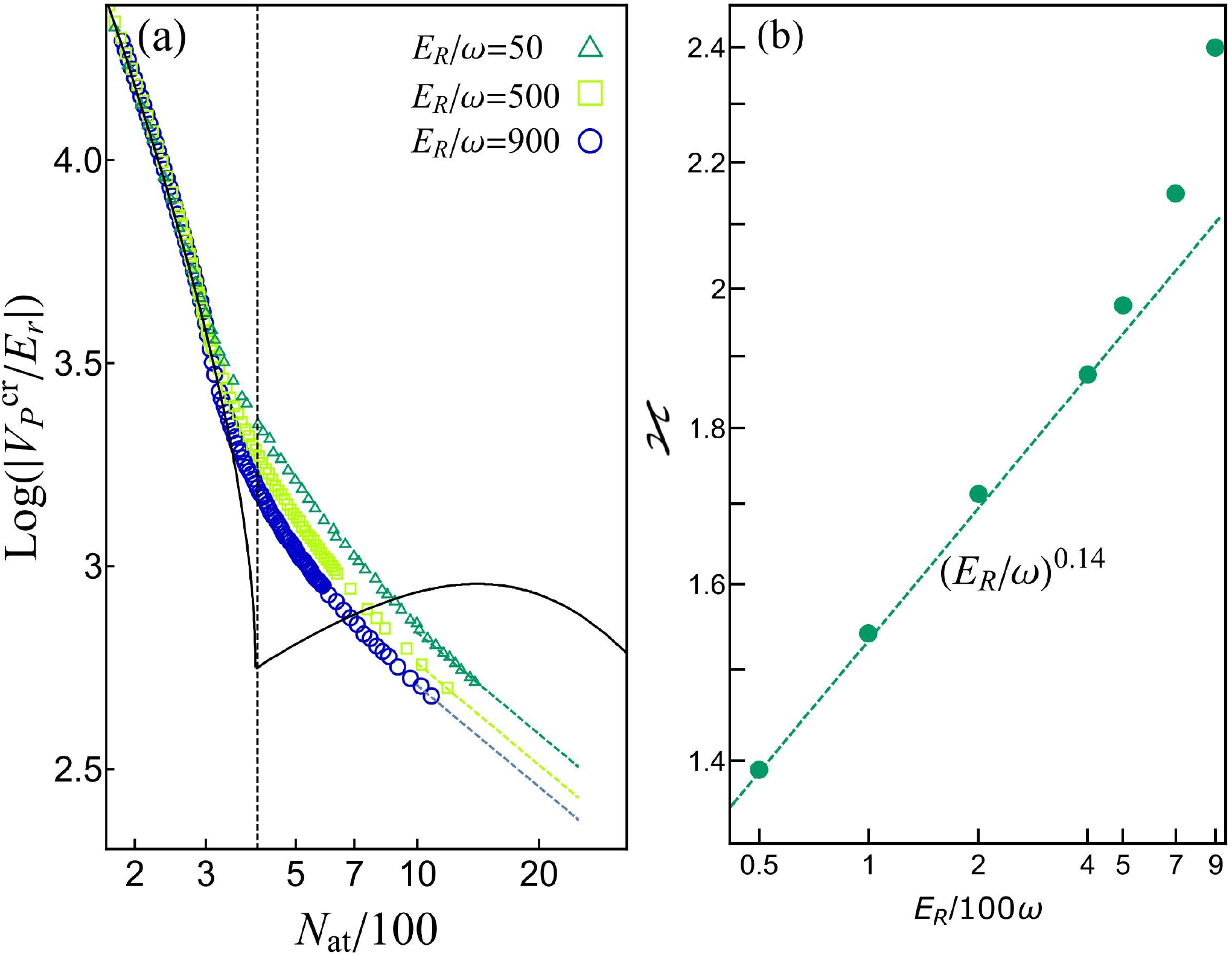}

\caption{ In (a), we show critical pumping strength as a function of particle number for different $\omega_z/E_R$. $\Delta_c$ is taken in the localized region. One can find when $E_R/\omega_z$ becomes bigger, the slope in nesting region increases. The gray solid line is the theoretical prediction that we expect as the $E_R/\omega_z\rightarrow\infty$ limit of the universal critical pumping strength curve. In (b), we show the slope as a function of $E_R/\omega_z$. The increase of the slope is found to be faster than power law increasing. As a result, we expect $\varkappa_{\rm Nest}\rightarrow\infty$ as $\omega_z/E_R\rightarrow 0$. }
\label{Nesting}
\end{figure}

Now we will present the assumptions of our effective theory. First of all, $\phi_{n_z}(z)$ is approximated by its large $n_z$ asymptotic expression $\phi_{n_z}(z)\sim \cos(\sqrt{(2n_z+1/2)\omega_z}z)$ for even $n_z$ and $\sin(\sqrt{(2n_z+1/2)\omega_z}z$ for odd $n_z$. Although the original approximation is only correct at large $n_z$ limit, here we take an approximation for every $n_z$. Further we drop the constant $1/2$ and employ $k_z^2=2n_z\omega_z$. Thus $\sum_{n_z}=\int \frac{|k_z|}{m\omega_z}dk_z$ and $\phi_{n_z}(z)=\cos (k_z z)$ or $\sin(k_z z)$. Second, in $x$ direction, if $\phi_{n_x}(x)$ is localized at $x_0$, then $\phi_{n_x}(x)=\delta(x-x_0)$. If the wave function is not localized, we apply a local density approximation and $\phi_{n_x}(x)$ is characterized by both position $x_0$ and wave vector $k_x$. The eigen-energy is $\varepsilon_{x_0,k_x}=2t\cos k_x+\frac{1}{2}\omega_x x_0^2$. Here we introduce $x_{\rm Mob}$ as the mobility edge in $x$ direction. For $x<x_{\rm Mob}$, the wave function $\phi_{n_x}(x)$ is extended and for $x>x_{\rm Mob}$, the wave function is localized. $\sum_{n_x}$ is approximated as a phase space integral $\int dk_x dx_0$. With all these approximations, Eq.~\ref{StaticStructure} and Eq.~\ref{Cr} can be calculated analytically and we find $V_{P}^{\rm cr}\propto N^{-1/3}_{\rm at}$ at large $N_{\rm at}$ limit irrelevant to the position of $x_{\rm Mob}$. The detailed calculation is given in supplementary material\cite{supp} and the theoretical prediction is shown in Fig.~\ref{Nesting}(a) in solid black curve. A kink at Fermi surface nesting atom number can be observed. 

Since the approximation in $z$ direction deviate from reality, we find the approximation is better when $E_R/\omega_z\gg 1$. Therefore we expect the critical pumping strength scaling of $N_{\rm at}$ in the middle range $N_{\rm at}$ will increase when $\omega_z/E_R$ is decreasing. Thus this prediction can be numerically checked by numerics. Interestingly, we find except for scaling around the Fermi surface nesting atom number, other scalings of critical pumping strength is invariant for different $\omega_z$. The scaling of critical pumping strength $|\log V_{P}^{\rm cr}/\log N_{\rm at}|$ is increasing when $\omega_z$ is decreasing. We are restricted by the system size and particle number in a numerical calculation, therefore we can only prove that the scaling is increasing when $\omega_z$ becomes smaller. A finite size scaling is carried out, and a divergent scaling is obtained in $\omega_z\rightarrow 0$ limit. All of the present data analysis can be equally done for experimental data. These signatures in critical pumping strength can then be identified as Fermi surface nesting effect.

{\color{blue}\emph{Conclusion}}  To summarize, we find the Fermi surface nesting effect in fermionic superradiant transition in a cavity can be verified with the help of trap induced localization. We find when the harmonic trap depth is effectively changed, there are two typical curves for superradiant transition critical pumping strength as a function of particle number. For shallow trap without localization effect, there are no signs of Fermi surface nesting, and for tight trap with localization effect, there is Fermi surface nesting signal which manifests as $\varkappa>1$ for $\varkappa=\log (V_{P}^{\rm cr})/\log N_{\rm at}$.  We also verified the tendency for $\varkappa\rightarrow \infty$ in zero trap frequency limit. We find the interplay between trap or localization and superradiance is quite interesting, and statistical effect can be magnified and thus benefits the generation of superradiance.

\color{blue}\emph{Acknowledgments.}\color{black}--- Y. C is supported by NSFC under Grant No. 12174358 and No. 11734010, and Beijing Natural Science Foundation (Z180013).

\newpage

\appendix

\setcounter{equation}{0}
\renewcommand{\theequation}{S\arabic{equation}}

\begin{widetext}
\newpage
\centerline{\bf Supplementary Material}

\section{Calculation of the off-diagonal matrix element $\langle n|\cos(k_0z)|n'\rangle$ in Eq.~\ref{StaticStructure}}
The transition matrix element  $\langle n|\cos(k_0z)|n'\rangle=\int dz \phi^*_{n}(z)\phi_{n'}(z)\cos (k_0z)$, $\phi_{n'}(z)$, where $\phi_{n(n')}(z)$ is the eigen-state of harmonic trap in z direction. $\hat{H}_{z}|n\rangle=(-\partial_z^2/2m+(m\omega_z z^2)/2)\phi_n(z)=\omega_n^z\phi_{n}(z)$, $\omega_n^z=(n+1/2)\omega_z$. $\hbar=1$ is employed. Here we introduce 
\ba
f_{nn'}=\langle n|e^{i k_0 z}|n'\rangle,
\ea
then we can find that
\ba
\int dz \phi^*_{n}(z)\phi_{n'}(z)\cos (k_0z)=\int dz \phi^*_{n}(z)\phi_{n'}(z){\rm Re}(e^{ ik_0z})={\rm Re}(\langle n|e^{ik_0z}|n'\rangle)= {\rm Re} (f_{nn'}).
\ea
Since the transition element is just the real part of $f_{nn'}$, therefore we will focus on calculation of $f_{nn'}$ in the following. Here we will present an algebraic method involving the annihilation and creation operators. Let us introduce $\hat{a}=\sqrt{\frac{m\omega_z}{2}}z+\sqrt{\frac{1}{2m\omega_z}}\partial_z$, $\hat{a}^\dag=\sqrt{\frac{m\omega_z}{2}}z-\sqrt{\frac{1}{2m\omega_z}}\partial_z$, then we have $z=\sqrt{\frac{1}{2m\omega_z}}(\hat{a}+\hat{a}^\dag)$, $\hat{H}_z=\omega_z(\hat{a}^\dag\hat{a}+\frac{1}{2})$. Then eigenstate $|n\rangle=(\hat{a}^\dag)^n/\sqrt{n!}|0\rangle$. Then we find
\ba
f_{nn'}=\langle n|e^{ik_0z}|n'\rangle=\frac{1}{\sqrt{n!n'!}}\langle0|\hat{a}^n e^{i\frac{k_0}{\sqrt{2m\omega_z}}(\hat{a}+\hat{a}^\dag)}(\hat{a}^{\dag})^{n'}|0\rangle
\ea
Let us introduce $\vartheta=k_0/\sqrt{2m\omega_z}$, then we have
\ba
f_{nn'}&=&\bra{n}e^{ik_0z}\ket{n'}=\frac{1}{\sqrt{n!n'!}}\langle0|\hat{a}^n e^{i\vartheta(\hat{a}+\hat{a}^\dag)}(\hat{a}^{\dag})^{n'}|0\rangle\nonumber\\
&=&\frac{1}{\sqrt{n!n'!}}\langle 0|e^{i\vartheta(\hat{a}+\hat{a}^\dag)}e^{-i\vartheta(\hat{a}+\hat{a}^\dag)}\hat{a}^ne^{i\vartheta(\hat{a}+\hat{a}^\dag)}(\hat{a}^{\dag})^{n'}|0\rangle.
\ea
Making use of Baker-Hausdorff formula, we find
\ba
&&e^{-i\vartheta(\hat{a}+\hat{a}^\dag)}\hat{a}^ne^{i\vartheta(\hat{a}+\hat{a}^\dag)}=(e^{-i\vartheta(\hat{a}+\hat{a}^\dag)}\hat{a}e^{i\vartheta(\hat{a}+\hat{a}^\dag)})^n\nonumber\\
&=&\left(\hat{a}-i\vartheta[\hat{a}+\hat{a}^\dag,\hat{a}]+\sum_{\ell=2}^\infty \frac{(-i\vartheta)^\ell}{\ell!}[\cdots,[\hat{a}+\hat{a}^\dag,\hat{a}]\cdots]\right)^n=(\hat{a}+i\vartheta)^n
\ea
Therefore
\ba
f_{nn'}&=&\frac{1}{\sqrt{n!n'!}}\langle 0|e^{i\vartheta(\hat{a}+\hat{a}^\dag)}(\hat{a}+i\vartheta)^n (\hat{a}^{\dag})^{n'}|0\rangle\nonumber\\
&=&\frac{1}{\sqrt{n!}}\langle 0|e^{i\vartheta(\hat{a}+\hat{a}^\dag)}\sum_\ell C_n^\ell (i\vartheta)^{n-\ell}\hat{a}^{\ell}|n'\rangle
\ea
Here we assume $n \leq n'$, then we have $\hat{a}^\ell |n'\rangle=(\sqrt{n'!/(n'-\ell)!})|n'-\ell\rangle$. Notice $e^{-i\vartheta (\hat{a}+\hat{a}^\dag)}|0\rangle$ is a coherent state, denoted as $|-i\vartheta\rangle$. $\langle 0|e^{i\vartheta (\hat{a}+\hat{a}^\dag)}=\langle -i\vartheta|$.
\ba
f_{nn'}&=&\frac{1}{\sqrt{n!}}\sum_\ell C_n^\ell(i\vartheta)^{n-\ell}\sqrt{\frac{n'!}{(n'-\ell)!}}\langle -i\vartheta|n-\ell\rangle\nonumber\\
&=&\frac{1}{\sqrt{n!}}\sum_\ell C_n^\ell(i\vartheta)^{n-\ell}\sqrt{\frac{n'!}{(n'-\ell)!}}\sqrt{\frac{1}{(n'-\ell)!}}\langle -i\vartheta|(\hat{a}^\dag)^{n'-\ell}|0\rangle\nonumber\\
&=&\frac{\sqrt{n'!}}{\sqrt{n!}}\sum_\ell C_n^\ell(i\vartheta)^{n-\ell}\frac{1}{(n'-\ell)!}((-i\vartheta)^*)^{n'-\ell}\langle -i\vartheta|0\rangle
\ea
In the last line of above equation, we have used $\langle-i\vartheta|\hat{a}^\dag=(-i\vartheta)^*\langle -i\vartheta|$. Finally, we have
\ba
f_{nn'}&=&\sqrt{n'!n!}\sum_{\ell=0}^{n}\frac{(i\vartheta)^{n-\ell}(i\vartheta)^{n'-\ell}}{\ell!(n-\ell)!(n'-\ell)!}\langle-i\vartheta|0\rangle
\ea
The factor $\langle -i\vartheta|0\rangle$ can be calculated as
\ba
\langle -i\vartheta|0\rangle=\langle0|e^{ik_0z}|0\rangle=\frac{1}{\sqrt{\pi m\omega_z}}\int dz e^{-m\omega_z z^2} e^{ik_0 z}=e^{-\frac{k_0^2}{4m\omega_z}}=e^{-\frac{1}{2}\vartheta^2}
\ea
The final result is
\ba
f_{nn'}=\sqrt{n!n'!}\sum_{\ell=0}^{n}\frac{(i\vartheta)^{n-\ell}(i\vartheta)^{n'-\ell}}{\ell!(n-\ell)!(n'-\ell)!}e^{-\frac{1}{2}\vartheta^2}
\ea
If $n'<n$, we find
\ba
f_{nn'}=\sqrt{n!n'!}\sum_{\ell=0}^{n'}\frac{(i\vartheta)^{n-\ell}(i\vartheta)^{n'-\ell}}{\ell!(n-\ell)!(n'-\ell)!}e^{-\frac{1}{2}\vartheta^2}
\ea
In Fig.~\ref{Matrix}, we show the matrix element of $|f_{nn'}|$ as a function of $n$ and $n'$. Meanwhile, in the maintext, we approximate $|n\rangle$ as  $\frac{1}{2}(|k_z\rangle\pm|-k_z\rangle)$, then $f_{nn'}=\frac{1}{2}(\delta_{k'_z,k_z+k_0}+\delta_{k'_z,k_z-k_0})$. We show the solid line as the delta function between $k_z$ and $k'_z$.

\begin{figure}[t]
\includegraphics[width=5cm]{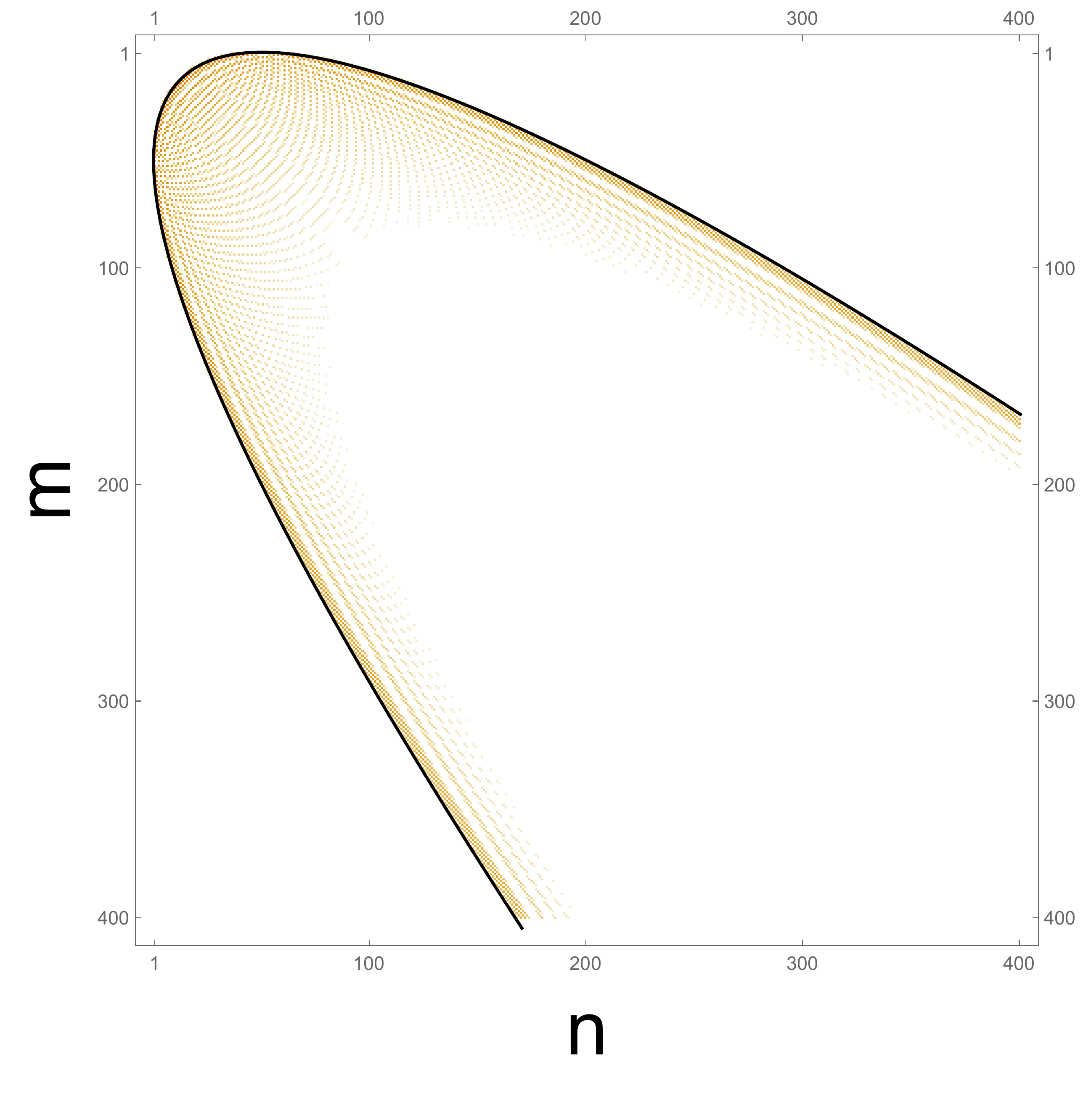}

\caption{Fixed $\vartheta=\sqrt{50}$, we show $|f_{mn}|^2$ more clearly by matrix density diagram, deep color mean nonzero value. As shown in the figure, when m and n is large, only a small number of points are non-zero, which behave like a dirac function. The black line is the result calculated by approximation ~\ref{S6} .}
\label{Matrix}
\end{figure}

\section{Predictions of the Effective theory}

Considering the fermions in a trap with optical lattice in x direction, the eigenstates are classified into two types. One type is itinerate wave-function and another is localized one. When the excitation energy is above the mobility edge, then the localized eigenstate is labeled by its position in x direction and $n_z$ in z direction.  The energy of eigenstate $|j,n_z\rangle$ is
\ba
\epsilon_{j,n_z}=\frac{1}{2}m\omega^2_x(ja_0)^2+n_z\omega_z,
\ea
where $a_0=\pi/k_0$ is the lattice length unit. On the other hand, if the eigen energy is smaller than the mobility edge, then we can approximate the excited state energy by local density approximation. $\epsilon_{j,k_x,n_z}=\frac{1}{2}m\omega^2_x(ja_0)^2+n_z\omega_z+2t\cos(k_x a_0)$, where $t$ is the hopping strength in x direction. From this dispersion relation, we can obtain the relation between chemical potential and the particle number $N_{\rm at}$. Here we are going to focus on the case when the localized states are the major states ( this is true when $N_{\rm at}$ is large ).
\ba
N_{\rm at}=\sum_{j,n_z}n_F(\epsilon_{j,n_z}-\mu)=\frac{2}{a_0}\int_0^{\sqrt{\frac{2\mu}{m\omega_x^2}}}dx\int_0^{\frac{\mu-\frac{1}{2}m\omega_x x^2}{\omega_z}} dn_z=\frac{4}{3\pi\omega_x\omega_z}\sqrt{\frac{2k_0^2}{m}}\mu^{3/2}
\ea
In above, we use$\sum_{j=-J}^{j=J}=\frac{1}{a_0}\int_{-Ja_0}^{Ja_0}dx$.Then we can see $\mu\propto N_{\rm at}^{2/3}$. More explicitly,
\ba
\mu=\left(\frac{3\pi\omega_x\omega_z}{8\sqrt{E_R}}N_{\rm at}\right)^{2/3}
\ea
Now we are going to check the critical pumping strength as a function of chemical potential $\mu$, then equivalently we get the critical pumping strength as a function of $N_{\rm at}$ from above relation. From Eq.~10, we know 
\ba
\left| \frac{V_P^{cr}(N_{at})}{E_R} \left|=\frac{-(\Delta_c'^2+\kappa^2)E_R}{4U_0\Delta_c'\chi(N_{\rm at})}\right.\right..
\ea
where the susceptibility $\chi$ is
\ba
\chi=\frac{1}{2\eta^2_0}\sum_{n,n'}\left| \int d \textbf{r}\phi^*_n(\textbf{r})\phi_{n'}(\textbf{r})\eta(\textbf{r})\right|^2\frac{n_F(E_n)-n_F(E_n')}{E_{n'}-E_n}
\ea
Attention that the quantum number $n$ represent all the quantum numbers of the eigenstate, including $j$, $k_x$ and $n_z$. Inspired by the asymptotic representation of hermite polynomials: $H_{2n_z}(z)=(-1)^{n_z}2^{n_z}(2n_z-1)!!e^{z^2/2}[\cos{(\sqrt{4n_z+1}z)}+O(\frac{1}{n_z^{1/4}})]$, $H_{2n_z+1}(z)=(-1)^{n_z}2^{n_z+1/2}(2n_z-1)!!e^{z^2/2}\sqrt{2n_z+1}[\sin{(\sqrt{4n_z+3}z)}+O(\frac{1}{n_z^{1/4}})]$, 
we can take following approximations for large $n_z$,
\ba
&&\langle z|2n_
z\rangle\approx\frac{1}{2\sqrt{L_{k_z}}}(\langle z|k_
z\rangle+\langle z|-k_
z\rangle)\theta(L_{k_z}^2-z^2),\nonumber\\&&\langle z|2n_
z+1\rangle\approx \frac{1}{2\sqrt{L_{k_z}}}(\langle z|k_
z\rangle-\langle z|-k_
z\rangle)\theta(L_{k_z}^2-z^2)(k_z>0) \label{S6}
\ea
where $|k_z\rangle$ is a momentum state $\langle z|k_z\rangle=e^{ik_zz}$,$L_{k_z}=\frac{\pi |k_z|}{4m\omega_z}$ and $k_z^2/2m=n_z\omega_z$. The range in z direction is due to the wave function is only obvious nonzero within the trap range.
Here the plus sign is for $n_z$ being even and the minus sign is for $n_z$ being odd. This approximation is good when $n_z$ is large. But we will take this approximation for all $n_z$. Under such approximation, we have
\ba
\langle 2n_z|e^{ik_0 z}|2n'_z\rangle&=&\frac{1}{4\sqrt{L_{k_z}L_{k_z'}}}\int_{-\rm{min}(L_{k_z},L_{k'_z})}^{\rm{min}(L_{k_z},L_{k'_z})} dz(\langle k_z|+\langle -k_z|)|z\rangle e^{ik_0 z}\langle z|(|k'_z\rangle+|-k'_z\rangle)
\ea
Let us denote $L_{z\rm min}={\rm min}(L_{k_z},L_{k'_z})$,$L_{z\rm max}={\rm max}(L_{k_z},L_{k'_z})$. Then we have
\ba
\langle 2n_z|e^{ik_0 z}|2n'_z\rangle&=&\frac{\sin((k_0+k'_z-k_z)L_{z\rm min})}{(k_0+k'_z-k_z)(2\sqrt{L_{z\rm max}L_{z\rm min}})}+\frac{\sin((k_0+k'_z+k_z)L_{z\rm min})}{(k_0+k'_z+k_z)(2\sqrt{L_{z\rm max}L_{z\rm min}})}+\nonumber\\
&&\frac{\sin((k_0-k'_z-k_z)L_{z\rm min})}{(k_0-k'_z-k_z)(2\sqrt{L_{z\rm max}L_{z\rm min}})}+\frac{\sin((k_0-k'_z+k_z)L_{z\rm min})}{(k_0-k'_z+k_z)(2\sqrt{L_{z\rm max}L_{z\rm min}})}
\ea
Notice that in $\omega_z\rightarrow 0^+$ limit, $L_{z\rm min}\rightarrow\infty$, such that $\sin( kL_{z\rm min})/k\approx\delta(k)$. It is easy to check that $\langle 2n_z|e^{-ik_0 z}|2n'_z\rangle=\langle 2n_z|e^{ik_0 z}|2n'_z\rangle$, $\langle 2n_z|\cos(k_0 z)|2n'_z+1\rangle=\langle 2n_z+1|\cos(k_0 z)|2n'_z\rangle=0$ and $\langle 2n_z+1|e^{ik_0z}|2n'_z+1\rangle=\langle 2n_z+1|e^{-ik_0z}|2n'_z+1\rangle$. What we really need is $|\langle 2n_z|e^{ik_0 z}|2n'_z\rangle|^2$, we have
\ba
\left|\langle 2n_z|\cos(k_0 z)|2n'_z\rangle\right|^2&=&(\langle 2n_z|e^{ik_0 z}|2n'_z\rangle)^2\nonumber\\
&\approx&\left(\frac{\sin((k_0+k'_z-k_z)L_{z\rm min})}{(k_0+k'_z-k_z)(2\sqrt{L_{z\rm max}L_{z\rm min}})}\right)^2+\left(\frac{\sin((k_0+k'_z+k_z)L_{z\rm min})}{(k_0+k'_z+k_z)(2\sqrt{L_{z\rm max}L_{z\rm min}})}\right)^2+\nonumber\\
&&\left(\frac{\sin((k_0-k'_z-k_z)L_{z\rm min})}{(k_0-k'_z-k_z)(2\sqrt{L_{z\rm max}L_{z\rm min}})}\right)^2+\left(\frac{\sin((k_0-k'_z+k_z)L_{z\rm min})}{(k_0-k'_z+k_z)(2\sqrt{L_{z\rm max}L_{z\rm min}})}\right)^2\nonumber\\
&\approx&\left(\delta_{k_z,k'_z+k_0}+\delta_{-k_z,k'_z+k_0}+\delta_{k_z,-k'_z+k_0}+\delta_{-k_z,k'_z+k_0}\right)/(4L_{z\rm max})
\ea
In the first $\approx$ in above equation, we dropped the cross terms. In the second $\approx$, we employed $\lim_{L_{z\rm min}\rightarrow\infty}(\sin (kL_{z\rm min})/k)^2\approx L_{z\rm min}\delta(k)$.
Meanwhile when we replace $n_z\omega_z$ by $k_z^2/2m$, the summation over $n_z$ is replaced by 
\ba
\sum_{n_z}=\int_0^{\infty} dk_z\frac{k_z}{m\omega_z}
\ea
By approximation of all the eigenstates are localized, the susceptibility can be written as
\ba
\chi(\mu)&&=\frac{1}{2}\sum_{j,n_z;j',n'_z}\frac{|\langle j, n_z|\cos(k_0z)\cos(k_0x)|j',n'_z\rangle|^2}{\epsilon_{j,n_z}-\epsilon_{j',n'_z}}(\theta(\mu-\epsilon_{j',n'_z})-\theta(\mu-\epsilon_{j,n_z}))\nonumber\\&&=\frac{1}{2}\sum_{\epsilon_{j',n'_z}<\mu,\epsilon_{j,n_z}>\mu}\frac{|\langle j, n_z|\cos(k_0z)\cos(k_0x)|j',n'_z\rangle|^2}{\epsilon_{j,n_z}-\epsilon_{j',n'_z}}+\frac{1}{2}\sum_{\epsilon_{j',n'_z}>\mu,\epsilon_{j,n_z}<\mu}\frac{|\langle j, n_z|\cos(k_0z)\cos(k_0x)|j',n'_z\rangle|^2}{\epsilon_{j,n_z}-\epsilon_{j',n'_z}}\nonumber\\&&=\sum_{\epsilon_{j',n'_z}>\mu,\epsilon_{j,n_z}<\mu}\frac{|\langle n_z|\cos(k_0z)|n'_z\rangle|^2|\langle j|\cos(k_0x)|j'\rangle|^2}{\epsilon_{j',n'_z}-\epsilon_{j,n_z}}\nonumber\\&&=\sum_{\epsilon_{j,n_z}<\mu;n'_z}\frac{|\langle n_z|\cos(k_0z)|n'_z\rangle|^2}{\epsilon_{j,n'_z}-\epsilon_{j,n_z}}\theta\left(\epsilon_{j,n'_z}-\mu\right)
\ea
Where $\ket{j}$ repersent the state localized in the jth site, now we introduce $x=ja_0$ Then $\sum_j=\frac{1}{a_0}\int dx$. In these approximations,
\ba
\chi(\mu)&&=\frac{1}{a_0}\int_{-\sqrt{\frac{2\mu}{m\omega_x^2}}}^{\sqrt{\frac{2\mu}{m\omega_x^2}}} dx\int^{\sqrt{2m(\mu-1/2m\omega_x^{2}x^{2})}}_{-\sqrt{2m(\mu-1/2m\omega_x^{2}x^{2})}}\frac{|k_z|dk_z}{2m\omega_z}\int \frac{|k^{\prime}_z|dk'_z}{2m\omega_z}\frac{|\langle n_z|\cos(k_0z)|n'_z\rangle|^2}{(n'_z-n_z)\omega_z}\theta\left(n'_z\omega_z+\frac{m\omega_xx^2}{2}-\mu\right)\nonumber\\&&=\frac{1}{a_0}\int_{-\sqrt{\frac{2\mu}{m\omega_x^2}}}^{\sqrt{\frac{2\mu}{m\omega_x^2}}} dx\int^{\sqrt{2m(\mu-1/2m\omega_x^2x^2)}}_{-\sqrt{2m(\mu-1/2m\omega_x^2x^2)}}\frac{dk_z|k_z|}{2m \omega_{z}}\int \frac{|k'_z|dk'_z}{2m\omega_z}\frac{(\delta_{k_z,k_z'-k_0}+\delta_{k_z,-k_z'-k_0}+\delta_{-k_z,k_z'-k_0}+\delta_{-k_z,-k_z'-k_0})}{4L_{z\rm max}(k^{\prime2}_z/2m-k^2_z/2m)}\nonumber
\\&&\theta\left(\frac{k_z^{\prime2}}{2m}+\frac{m\omega^2_x x^2}{2}-\mu\right)\nonumber
\\&&=\frac{1}{a_0}\int_{-\sqrt{\frac{2\mu}{m\omega_x^2}}}^{\sqrt{\frac{2\mu}{m\omega_x^2}}} dx\int^{\sqrt{2m(\mu-1/2m\omega_x^2x^2)}}_{-\sqrt{2m(\mu-1/2m\omega_x^2x^2)}}\frac{|k_z|dk_z}{2m\omega_z}\int\frac{|k^{\prime}_z|dk^{\prime}_z}{2\pi\rm{max}(|k_z|,|k^{\prime}_z|)}\nonumber\\
&&\frac{(\delta_{k_z,k_z'-k_0}+\delta_{k_z,-k_z'-k_0}+\delta_{-k_z,k_z'-k_0}+\delta_{-k_z,-k_z'-k_0})}{(k^{\prime 2}_z/2m-k^2_z/2m)}\theta\left(\frac{k^{\prime2}_z}{2m}+\frac{m\omega^2_xx^2}{2}-\mu\right)\nonumber
\\&&=\frac{1}{a_0}\int_{-\sqrt{\frac{2\mu}{m\omega_x^2}}}^{\sqrt{\frac{2\mu}{m\omega_x^2}}} dx\int^{\sqrt{2m(\mu-1/2m\omega_x^2x^2)}}_{-\sqrt{2m(\mu-1/2m\omega_x^2x^2)}}\frac{|k_z|dk_z}{2m\omega_z}\frac{|k_z+k_0|}{2\pi\rm{max}(|k_z|,|k_z+k_0|)}\frac{\theta\left(\frac{(k_z+k_0)^2}{2m}+\frac{m\omega^2_xx^2}{2}-\mu\right)}{(k_z+k_0)^2/2m-k_z^2/2m}
\nonumber
\\&&+\frac{1}{a_0}\int_{-\sqrt{\frac{2\mu}{m\omega_x^2}}}^{\sqrt{\frac{2\mu}{m\omega_x^2}}} dx\int^{\sqrt{2m(\mu-1/2m\omega_x^2x^2)}}_{-\sqrt{2m(\mu-1/2m\omega_x^2x^2)}}\frac{|k_z|dk_z}{2m\omega_z}\frac{|k_z+k_0|}{2\pi\rm{max}(|k_z|,|k_z+k_0|)}\frac{\theta\left(\frac{(k_z+k_0)^2}{2m}+\frac{m\omega^2_xx^2}{2}-\mu\right)}{(k_z+k_0)^2/2m-k_z^2/2m}\nonumber
\\&&+\frac{1}{a_0}\int_{-\sqrt{\frac{2\mu}{m\omega_x^2}}}^{\sqrt{\frac{2\mu}{m\omega_x^2}}} dx\int^{\sqrt{2m(\mu-1/2m\omega_x^2x^2)}}_{-\sqrt{2m(\mu-1/2m\omega_x^2x^2)}}\frac{|k_z|dk_z}{2m\omega_z}\frac{|k_z-k_0|}{2\pi\rm{max}(|k_z|,|k_z-k_0|)}\frac{\theta\left(\frac{(k_z-k_0)^2}{2m}+\frac{m\omega^2_xx^2}{2}-\mu\right)}{(k_z-k_0)^2/2m-k_z^2/2m}\nonumber\\&&+\frac{1}{a_0}\int_{-\sqrt{\frac{2\mu}{m\omega_x^2}}}^{\sqrt{\frac{2\mu}{m\omega_x^2}}} dx\int^{\sqrt{2m(\mu-1/2m\omega_x^2x^2)}}_{-\sqrt{2m(\mu-1/2m\omega_x^2x^2)}}\frac{|k_z|dk_z}{2m\omega_z}\frac{|k_z-k_0|}{2\pi\rm{max}(|k_z|,|k_z-k_0|)}\frac{\theta\left(\frac{(k_z-k_0)^2}{2m}+\frac{m\omega^2_xx^2}{2}-\mu\right)}{(k_z-k_0)^2/2m-k_z^2/2m}\nonumber
\ea
\ba
&&=\frac{2}{a_0}\int_{-\sqrt{\frac{2\mu}{m\omega_x^2}}}^{\sqrt{\frac{2\mu}{m\omega_x^2}}} dx\int^{\sqrt{2m(\mu-1/2m\omega_x^2x^2)}}_{-\sqrt{2m(\mu-1/2m\omega_x^2x^2)}}\frac{|k_z|dk_z}{2m\omega_z}\frac{|k_z+k_0|}{2\pi\rm{max}(|k_z|,|k_z+k_0|)}\frac{\theta\left(\frac{(k_z+k_0)^2}{2m}+\frac{m\omega^2_xx^2}{2}-\mu\right)}{(k_z+k_0)^2/2m-k_z^2/2m}
\nonumber
\\&&+\frac{2}{a_0}\int_{-\sqrt{\frac{2\mu}{m\omega_x^2}}}^{\sqrt{\frac{2\mu}{m\omega_x^2}}} dx\int^{\sqrt{2m(\mu-1/2m\omega_x^2x^2)}}_{-\sqrt{2m(\mu-1/2m\omega_x^2x^2)}}\frac{|k_z|dk_z}{2m\omega_z}\frac{|k_z-k_0|}{2\pi\rm{max}(|k_z|,|k_z-k_0|)}\frac{\theta\left(\frac{(k_z-k_0)^2}{2m}+\frac{m\omega^2_xx^2}{2}-\mu\right)}{(k_z-k_0)^2/2m-k_z^2/2m}\nonumber
\ea

Noticed that when we change $k_z$ to $-k_z$, the intergred function doesn't change, so we just need to consider $k_z>0$.
\ba
\chi(\mu)=&&\frac{8 }{a_0}\int_{0}^{\sqrt{\frac{2\mu}{m\omega_x^2}}} dx\int^{\sqrt{2m(\mu-1/2m\omega_x^2x^2)}}_{0}\frac{k_zdk_z}{2m\omega_z}\frac{k_z+k_0}{2\pi\rm{max}(k_z,k_z+k_0)}\frac{\theta\left(\frac{(k_z+k_0)^2}{2m}+\frac{m\omega^2_xx^2}{2}-\mu\right)}{(k_z+k_0)^2/2m-k_z^2/2m}\nonumber
\\&&+\frac{8 }{a_0}\int_{0}^{\sqrt{\frac{2\mu}{m\omega_x^2}}} dx\int^{\sqrt{2m(\mu-1/2m\omega_x^2x^2)}}_{0}\frac{k_zdk_z}{2m\omega_z}\frac{|k_z-k_0|}{2\pi\rm{max}(k_z,|k_z-k_0|)}\frac{\theta\left(\frac{(k_z-k_0)^2}{2m}+\frac{m\omega^2_xx^2}{2}-\mu\right)}{(k_z-k_0)^2/2m-k_z^2/2m}\nonumber
\ea

The calculation is carried in the three cases:$\mu/E_R<1/4,1/4<\mu/E_R<1,\mu/E_R>>1$.

For $\mu<1/4E_R$, the theta function in $\chi$ is always satisfied. 
\begin{equation}
    \begin{split}
        \chi(\mu)=\frac{4\sqrt{\mu E_R}}{\pi^2\omega_z\omega_x}\left(-1+1\sqrt{-1+\frac{E_R}{4\mu}}\arctan(1/\sqrt{-1+\frac{E_R}{4\mu}})\right) 
    \end{split}
\end{equation}

For $1/4<\mu/E_R<1$
\begin{equation}
\begin{split}
    \chi(\mu)&=\frac{-4\mu}{\pi^2 \omega_x\omega_z}\left(0.5\sin\left(2\arcsin\sqrt{1-\frac{E_R}{4\mu}}\right)+\arcsin\sqrt{1-\frac{E_R}{4\mu}}\right)+\frac{4 }{\pi^2 \omega_x\omega_z}\sqrt{E_R\mu}\sqrt{1-\frac{E_R}{4\mu}}\\
    &+\frac{2 \sqrt{E_R\mu}}{\omega_x\omega_z\pi^2}\left(-2-\sqrt{1-\frac{E_R}{4\mu}}\log\left(\frac{1-\sqrt{1-\frac{E_R}{4\mu}}}{1+\sqrt{1-\frac{E_R}{4\mu}}}\right)\right)
    \end{split}
\end{equation}

For $\mu>>E_R$
\begin{equation}
    \begin{split}
        \chi(\mu)&\approx\frac{4}{\pi^2\omega_x\omega_z}\sqrt{\mu E_R}
        \end{split}
\end{equation}
Apply critical condition:$\Delta'_c+4\chi \Delta^{\prime 2}_c/(\Delta^{\prime 2}_c+\kappa^2)=0$, we get:
\begin{equation}
|V_P^{cr}/E_R|\approx\frac{\Delta^{\prime 2}_c+\kappa^2}{16\Delta'_cU_0}(\pi^2\omega_x\omega_z/E^2_R)(\frac{3\pi\omega_x\omega_z}{8E^2_R}N_{at})^{-\frac{1}{3}}\propto N_{at}^{-1/3}    
\end{equation}
\begin{figure}[h]
\includegraphics[width=7cm]{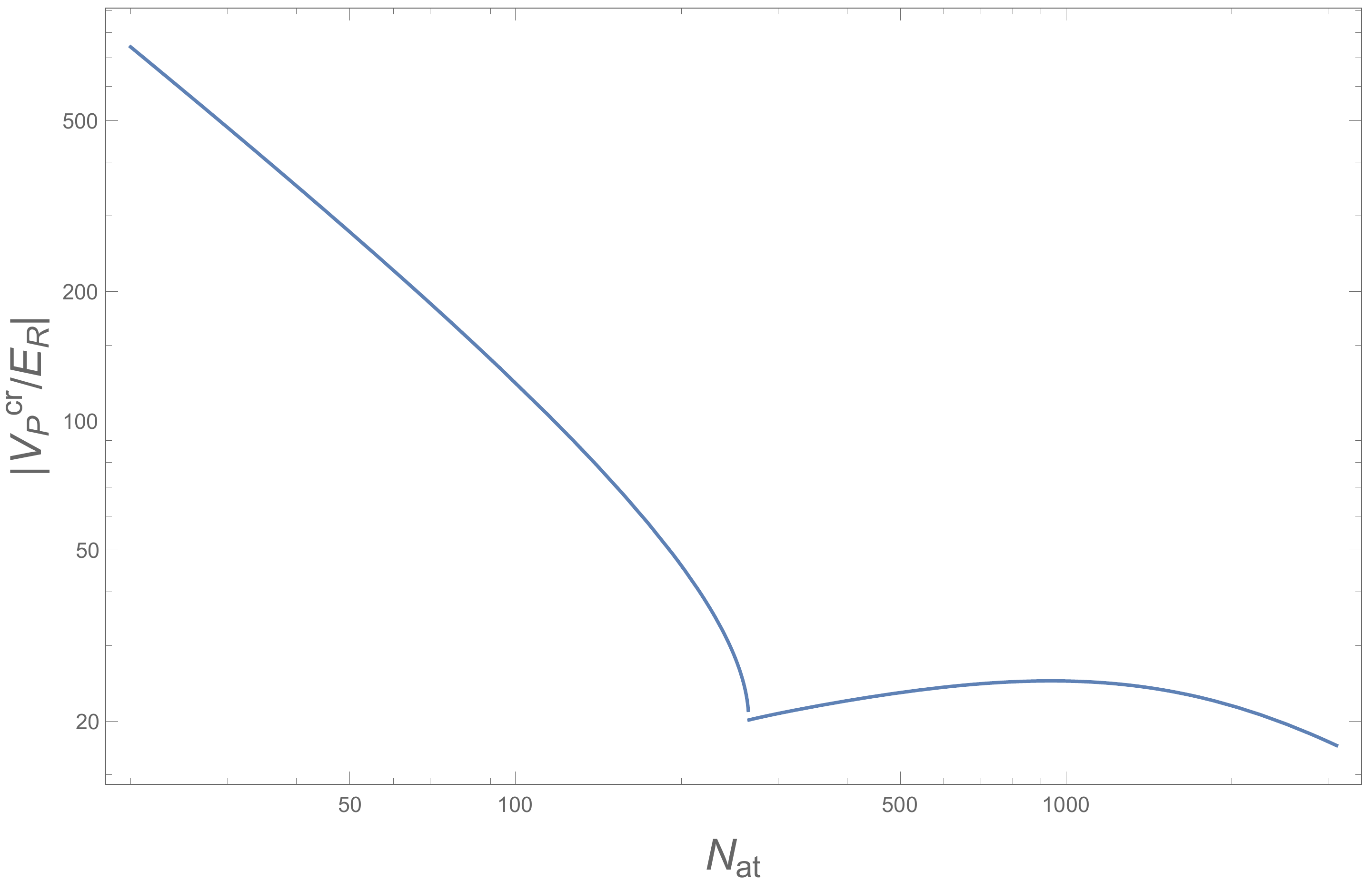}

\caption{We show $|V^{cr}_P(N_{at})/E_R|$ more clearly by diagram, from above calculation, we can see that the shape of this universal curve is independent of $\omega_z,\omega_x,\Delta'_c$}
\label{jiexi}
\end{figure}

\end{widetext}


\begin{thebibliography}{99}

\bibitem{Esslinger07} 
F. Brennecke, T. Donner, S. Ritter, T. Bourdel, M. K{\"o}hl, and T. Esslinger, \emph{Nature} (London) {\bf 450}, 268 (2007).

\bibitem{Colombe07} 
Y. Colombe, T. Steinmetz, G. Dubios, F. Linke, D. Hunger, and J. Reichel, \emph{Nature} (London) {\bf 450}, 272 (2007).

\bibitem{Esslinger10}
K. Baumann, C. Guerlin, F. Brennecke, and T. Esslinger, \emph{Nature} (London) {\bf 464}, 1301 (2010).



\bibitem{Dicke54}
R. H. Dicke, \emph{Phys. Rev.} {\bf 93}, 99 (1954).

\bibitem{Lieb73}
K. Hepp, and E. H. Lieb, \emph{Ann. Phys.} (N. Y.) {\bf 76}, 360 (1973).




\bibitem{Esslinger11}
K. Baumann, R. Mottl, F. Brennecke, and T. Esslinger, \emph{Phys. Rev. Lett.} {\bf 107}, 140402 (2011).

\bibitem{Roton12}
R. Mottl, F. Brennecke, K. Baumann, R. Landig, T. Donner, and T. Esslinger, \emph{Science} {\bf 336}, 1570–1573 (2012).

\bibitem{Esslinger13}
F. Brennecke, R. Mottl, K. Baumann, R. Landig, T. Donner, and T. Esslinger, \emph{Proceedings of the National Academy of Sciences} {\bf 110},  11763–11767 (2013).

\bibitem{Hemmerich14}
J. Klinder, H. Ke$\beta$ler, M. Wolke, L. Mathey, and A. Hemmerich, \emph{Proceedings of the National Academy of Sciences} {\bf 112}, 3290-3295 (2014).

\bibitem{Esslinger15}
R. Landig, F. Brennecke, R. Mottl, T. Donner, and T. Esslinger, \emph{Nature Commun.} {\bf 6}, 7046 (2015). 


\bibitem{Hemmerich15ex} 
J. Klinder, H. Ke\ss ler, M. R. Bakhtiari, M. Thorwart, and A. Hemmerich, 
\emph{Phys. Rev. Lett.} {\bf 115}, 230403 (2015).

\bibitem{Esslinger16}
R. Landig, L. Hruby, N. Dogra, M. Landini, R. Mottl, T. Donner, and T. Esslinger, 
\emph{Nature} {\bf 532}, 476 (2016).

\bibitem{Esslinger18} 
L. Hruby, N. Dogra, M. Landini, T. Donner, and T. Esslinger, 
\emph{PNAS} {\bf 115}, 3279 (2018).

\bibitem{Esslinger20}
X. Li, D. Dreon, P. Zupancic, A. Baumgartner, A. Morales, W. Zheng, N. Cooper, T. Donner, and T. Esslinger, 
\emph{Phys. Rev. Res.} {\bf 3}, L012024 (2021).

\bibitem{ContinousSS17}
J. Léonard, A. Morales, P. Zupancic, T. Esslinger and T. Donner \emph{Nature} {\bf543} 87-90 (2017).












\bibitem{EsslingerSS17}
J. L\'{e}onard, A. Morales, P. Zupancic, T. Donner, and T. Esslinger, 
Science {\bf 358} 1415–1418 (2017).

\bibitem{EsslingerSS18}
A. Morales, P. Zupancic, J. L\'{e}onard, T. Esslinger, and T. Donner, 
Nature Materials {\bf 17} 686–690 (2018).

\bibitem{Review21}
F. Mivehvar and F. Piazza and T. Donner and H. Ritsch, Annals of Physics, {\bf 70}, 1 (2021).


\bibitem{Simons14}
J. Keeling, M.J. Bhaseen, and B.D. Simons, \emph{Phys. Rev. Lett.} {\bf112}, 143002 (2014).

\bibitem{Piazza14}
F. Piazza and P. Strack, \emph{Phys. Rev. Lett.} {\bf112}, 143003
(2014).

\bibitem{Yu14}
Y. Chen, Z. Yu, and H. Zhai, \emph{Phys. Rev. Lett.} {\bf112}, 143004 (2014).

\bibitem{Yu15}
Y. Chen, H. Zhai, and Z. Yu, \emph{Phys. Rev. A} {\bf91}, 021602(R) (2015).

\bibitem{Wu21}
X. Zhang, Y. Chen, Z. Wu, J. Wang, J. Fan, S. Deng, and H. Wu, \emph{Science}, {\bf 373}, 1359–1362 (2021).

\bibitem{Ketterle21}
Y. Margalit, Y.-K. Lu, F. C. Top, and W. Ketterle, \emph{Science}, {\bf 374}, 976-979 (2021). 

\bibitem{Deb21}
A. B. Deb, and N. Kj\ae rgaard, \emph{Science}, {\bf 374}, 972-975 (2021). 

\bibitem{YeJ21}
C. Sanner, L. Sonderhouse, R. B. Hutson, L. Yan, W. R. Milner, and J. Ye, \emph{Science}, {\bf 374}, 979-983 (2021).


\bibitem{supp}
See supplementary material for the details. In the supplementary material, we include the calculations for the off-diagonal matrix element $\langle n|\cos(k_0z)|n'\rangle$, and the results for an approximate field theory.


\end{thebibliography}
\end{document}